\documentstyle[prb,aps,amssymb]{revtex}

\begin{document}
\draft
\twocolumn[\hsize\textwidth\columnwidth\hsize\csname@twocolumnfalse%
\endcsname

\title{Surface Properties of Aperiodic Ising Quantum Chains}
\author{Joachim Hermisson}
\address{Institut f\"{u}r Theoretische Physik,  
         Universit\"{a}t T\"{u}bingen, Auf der Morgenstelle 14,
         D-72076 T\"{u}bingen, Germany}
\author{Uwe Grimm}
\address{Institut f\"{u}r Physik,
         Technische Universit\"{a}t Chemnitz,
         D-09107 Chemnitz, Germany}
\date{Phys.~Rev.~\bf{B 57} 673-676 (1 January 1998) }
\maketitle

\begin{abstract}We consider Ising quantum chains with a quenched 
aperiodic disorder of the coupling constants given through general 
substitution rules. The critical scaling behavior of several bulk 
and surface quantities is obtained by exact real space renormalization.   
\end{abstract}

, ]
\narrowtext

\paragraph*{Introduction.}
Ising quantum chains with aperiodically modulated coupling constants
have been the topic of several recent 
publications,\cite{TIB,IT94,BBT,IT,HGB,ITKS} see also papers cited in 
Ref.~\onlinecite{GB96}. These systems describe 2d classical Ising
models, with couplings varying along layers, in the Hamiltonian limit and
provide a non-trivial, but nevertheless easily accessible class of
models to study the effect of aperiodic (dis)order on
thermodynamic properties and critical behavior. Scaling
arguments, put forward by Luck, lead to the following predictions 
on the relevance of aperiodic modulations:\cite{L} 
The critical behavior is of
Onsager-type as long as the fluctuations of the couplings are bounded,
but resembles the randomly disordered case for unbounded
fluctuations. Of particular interest is the marginal case with
(in terms of the system size) logarithmically diverging fluctuations. Here,
a non-universal scaling behavior is predicted 
with critical exponents that depend continuously on the coupling constants.

For sequences of coupling constants obtained from substitution rules,
we demonstrated recently,\cite{HGB} how Luck's criterion can be 
derived within an exact renormalization scheme, using a decimation 
procedure proposed in Ref.~\onlinecite{IT}. This way, an exact expression 
for the scaling exponent $z$ of the mass gap at criticality was
obtained.

In this article, we show how the renormalization approach can be
extended to describe the fermionic eigenvectors of the quantum
chains. This leads to the exact determination of the critical scaling 
behavior of several bulk and surface quantities, like the bulk
energy density or the surface magnetization. Quite recently, a similar
derivation (using the decimation of Ref.~\onlinecite{IT}) 
has been given for some special examples. \cite{ITKS} 
These results are contained as special cases here.

\paragraph*{Ising quantum chains and substitution sequences.}
The Ising quantum chain is given by the Hamiltonian 
\begin{equation}
\bbox{H}_{N}^{} = -\frac{1}{2}\left(\sum_{k=1}^{N} 
     \varepsilon_{k}^{}\,\bbox{\sigma}^{x}_{k}\bbox{\sigma}^{x}_{k+1} + 
     \sum_{k=1}^{N} h_k^{} \bbox{\sigma}^{z}_{k}\right)
\label{eq:QC}
\end{equation}
with Pauli matrices $\bbox{\sigma}^{x,z}_{k}$ acting
on the $k$\/th site, and site-dependent transverse magnetic fields $h_k$. 
The site-dependent coupling constants $\varepsilon_k$ shall be drawn
from a set of values $\varepsilon_{a_i}$, where the label $a_i$
is taken from an $n$-letter alphabet ${\cal A}=\{a_1,a_2,\ldots,a_n\}$.
Initially, the field variables are chosen as $h_k\equiv 1$;
but later on, as parameters in the renormalization transformation (RT), 
they become site-dependent and will correspondingly 
be labeled as $h_{k+1}=h_{a_i}$ (if $\varepsilon_{k}=\varepsilon_{a_i}$).

We choose the couplings according to sequences generated by iterated
application of a substitution rule \mbox{$\varrho: a_i \to w(a_i)$} on 
${\cal A}$. Some important properties of the substitution
chain are already contained in the corresponding 
$n\times n$ substitution matrix
$\bbox{M}_\varrho$, whose elements
$(\bbox{M}_\varrho)_{ij}^{}$ give the number of letters $a_j$
contained in the words $w(a_i)$.  
So, the asymptotic densities of the letters are given by the entries
of the (statistically normalized) right eigenvector to the
Perron-Frobenius (PF) eigenvalue $\lambda_1= \lambda_{\text{PF}}$
which is just the rescaling
factor of the chain length. The fluctuations of the cumulated
deviation of the couplings from the mean coupling grow as a power of
the chain length, i.e.\ as $N^\omega$, with a {\em
wandering exponent} $\omega$ determined by the two largest
eigenvalues (in modulus)
\begin{equation}
\omega = \frac{\log |\lambda_2|}{\log \lambda_{\text{PF}}} \; .
\end{equation}  
Thus fluctuations diverge for $|\lambda_2| \ge 1$, whereas the marginal
case is connected to $|\lambda_2|=1$.

The Ising quantum chain is critical, in the sense that the mass gap
vanishes in the thermodynamic limit, if the geometric mean of the
reduced couplings $\varepsilon_j/h_j$ approaches 1 for $N\to\infty$. 
For couplings determined by substitution sequences, this is equivalent to the
condition that the vector $\bbox{y}$ of logarithms of the reduced couplings 
\begin{equation}\label{crit}
\bbox{y}_i = \log \left(\frac{h^2_{a_i}}{\varepsilon^2_{a_i}}\right)
\qquad (i=1,2,\ldots,n)
\end{equation}
is perpendicular to the PF eigenvector of $\bbox{M}_\varrho$.

For a general set of coupling constants $\varepsilon_{k}^{}$,
the quantum chain can be written as a free-fermion model via a Jordan-Wigner
transformation and diagonalized canonically, resulting in \cite{LSM}
\begin{equation} \label{ffh}
\bbox{H}_N^{}=\sum^N_{q=1}\Lambda_q^{} 
(\bbox{\eta}_q^\dagger\bbox{\eta}_q^{} - 
{\case{1}{2}\bbox{\openone})} + C_N\bbox{\openone}
\end{equation}
where $\bbox{\eta}_q^\dagger$ and $\bbox{\eta}_q^{}$ are $N$ 
fermionic creation and annihilation operators and where
$C_N$ is some constant. The dimensionless excitation 
energies $\Lambda_q^{}$ (which can be ordered as
$0 \le \Lambda_1 \le \Lambda_2 \le \dots \le \Lambda_N$)
satisfy the linear difference equations
\begin{mathletters}\label{fermeq}\begin{eqnarray}
\Lambda_q^{} \bbox{\psi}_k^{(q)} &=& -\, h_k^{} \bbox{\phi}_k^{(q)} - 
\varepsilon_k^{} \bbox{\phi}_{k+1}^{(q)} \; , \\
\Lambda_q^{} \bbox{\phi}_k^{(q)} &=& -\, \varepsilon_{k-1}^{} 
\bbox{\psi}_{k-1}^{(q)} - h_k^{} \bbox{\psi}_k^{(q)} \; .
\end{eqnarray}\end{mathletters}%
This corresponds to an eigenvalue problem of the matrix
\begin{equation}  \label{fh}
\bbox{\cal H} = \left(
\begin{array}{@{}cccccccc@{}}
0&h_{\scriptscriptstyle 1}^{}&0&0&0&\cdots&0&
\varepsilon_{\scriptscriptstyle N}^{}\\
h_{\scriptscriptstyle 1}^{}&0&
\varepsilon_{\scriptscriptstyle 1}^{}&0&0&\cdots&0&0\\
0&\varepsilon_{\scriptscriptstyle 1}^{}&0&
h_{\scriptscriptstyle 2}^{}&0&\cdots&0&0\\
0&0&h_{\scriptscriptstyle 2}^{}&0&
\varepsilon_{\scriptscriptstyle 2}^{}&\cdots&0&0\\
\vdots&\vdots&\ddots&\ddots&\ddots&\ddots&&\vdots\\
0&0&\cdots&0&h_{\scriptscriptstyle N-1}^{}&
0&\varepsilon_{\scriptscriptstyle N-1}^{}&0\\
0&0&\cdots&0&0&\varepsilon_{\scriptscriptstyle N-1}^{} 
&0&h_{\scriptscriptstyle N}^{}\\
\varepsilon_{\scriptscriptstyle N}^{}&0&\cdots&0&0&0&
h_{\scriptscriptstyle N}^{}&0
\end{array}\right)
\end{equation}
where $\varepsilon_N^{}\equiv 0$ for
{\em free}\/ boundary conditions which are employed in the 
study of surface properties.  Here,
$\bbox{\psi}_k^{(q)}$ and $\bbox{\phi}_k^{(q)}$ are the even and 
the odd components of the eigenvector connected to the $q$\/th eigenvalue,
respectively.

\paragraph*{Renormalization scheme for fermionic eigenvectors.}
Following the presentation in Ref.~\onlinecite{HGB}, we now introduce the
renormalization technique based on $\bbox{\cal S}$-transfer matrices
and their star-products.\cite{RED} For technical reasons, we restrict
the discussion to substitution rules of the form 
\begin{equation} \label{subs}
\varrho: a_i \to a_i w_i
\end{equation}
where $w_i$ are $n$ arbitrary words in the $n$-letter alphabet ${\cal A}$,
and only comment on the extension to the general case. We introduce 
$\bbox{\cal S}$-matrices that transform the components of the 
eigenvectors of Eq.~(\ref{fh}) as
\begin{equation}
\left(\begin{array}{@{}c@{}} \bbox{\psi}_{k}\\ 
      \bbox{\phi}_{l+1}\end{array} \right) =
 \bbox{\cal S}_{k|l}
\left(\begin{array}{@{}c@{}} \bbox{\phi}_{k+1}\\ 
      \bbox{\psi}_{l}\end{array} \right)
\end{equation}
with
\begin{equation}
\bbox{\cal S}_{k|l} =
\bbox{\cal S}_{k|k+1} \ast \bbox{\cal S}_{k+1|k+2}
\ast \ldots \ast \bbox{\cal S}_{l-1|l} \; ,
\end{equation} 
where the \mbox{$\ast$-product} of two $2\times2$ matrices is defined as
\begin{equation}\label{star}
\left(\begin{array}{@{}c@{\;\;}c@{}}
e&\bar{e}\\r&\rho
\end{array} \right) \ast
\left(\begin{array}{@{}c@{\;\,}c@{}}
\rho_1^{}&r_1^{}\\\bar{o}&o
\end{array} \right) = 
\left(\begin{array}{@{}c@{\;\;}c@{}} e&0\\0&o \end{array} \right)+
\frac{1}{1-\rho \rho_1^{}}
\left(\begin{array}{@{}c@{\;\;}c@{}}
\bar{e} r \rho_1^{}&\bar{e} r_1^{}\\
\bar{o} r &\bar{o}r_1^{} \rho
\end{array} \right) \, .
\end{equation}
According to Eq.~(\ref{fh}), the elementary $\bbox{\cal S}$-matrices are
\begin{equation}
\label{sj}
\bbox{\cal S}_{k|k+1} = \left(
\begin{array}{@{\,}cc@{\,}}
\varepsilon_{k}^{-1} \Lambda & -\varepsilon_{k}^{-1} h_{k+1}^{} \\[1mm]
- \varepsilon_{k+1}^{-1} h_{k+1}^{} & \varepsilon_{k+1}^{-1} \Lambda 
\end{array} \right) 
\end{equation}
where $\Lambda$ denotes an eigenvalue of $\bbox{\cal H}$.
In order to establish the RT, the fields and 
vector components are labeled according to the locally varying
coupling constants. Furthermore, two additional asymmetry
coefficients $\kappa_{a_i}^\pm$ have to be introduced 
for every letter $a_i \in {\cal A}$.
This way, we obtain $n^2$ different elementary 
$\bbox{\cal S}$-transfer matrices
through
\begin{equation}
\left(\begin{array}{@{}c@{}} \bbox{\psi}_{a_i,k}\\ 
\bbox{\phi}_{a_j,k+2}\end{array} \right) =
 \bbox{\cal S}_{a_i|a_j}
\left(\begin{array}{@{}c@{}} \bbox{\phi}_{a_i,k+1}\\
\bbox{\psi}_{a_j,k+1}\end{array} \right)
\end{equation}
where the second label refers to the position, and where
\begin{equation}
\bbox{\cal S}_{a_i|a_j} = \left(
\begin{array}{@{\,}cc@{\,}} \varepsilon_{a_i}^{-1} \kappa_{a_i}^+ \Lambda
& -\varepsilon_{a_i}^{-1} h_{a_i}^{}\\[1mm] 
-\varepsilon_{a_j}^{-1}h_{a_i}^{} & \varepsilon_{a_j}^{-1}
\kappa_{a_i}^- \Lambda \end{array} \right) \; .
\end{equation}
Unlike the fields or the couplings, the eigenvector components depend not
only through the $a_i$, but also {\em explicitly}\/ on the position, 
so we have to keep the labels here.
The RT now reverses the substitution steps by building 
\mbox{$\ast$-products} of $\bbox{\cal S}$-transfer matrices
corresponding to the words $w_i$ given by the substitution rule.
The resulting RT's of $\kappa_{a_i}^\pm \Lambda$ and $h_{a_i}$ have been
described in Ref.~\onlinecite{HGB}. 
For coupling constants fulfilling the 
criticality condition, the vector $\bbox{y}$ of logarithms of the 
reduced couplings is found to scale with the next-to-leading 
eigenvalue $\lambda_2$ of $\bbox{M}_\varrho$.
Especially, in the marginal case with $\lambda_2 = 1$, $\bbox{y}$
converges to an eigenvector corresponding to an eigenvalue $1$ of 
the {\em transpose}\/ $\bbox{M}_{\varrho}^t$ 
of the substitution matrix. 

Let us now concentrate on the RT equations for the vector components of
$\bbox{\psi}$ and $\bbox{\phi}$. 
Since iterated $\ast$-multi\-plications
of the $\bbox{\cal S}$-matrices 
act as a pure decimation on the vector components, the
remaining components of the renormalized vectors
$\bbox{\tilde{\psi}}$ and $\bbox{\tilde{\phi}}$ differ from
the original ones only through relabeling and a normalization
factor
\begin{equation}
\bbox{\tilde{\psi}}_m  =  C_\psi \,\bbox{\psi}_k \; ,\quad 
\bbox{\tilde{\phi}}_{m+1} = C_\phi \,\bbox{\phi}_{k+1} \; ,
\label{cc}
\end{equation}
with $k/m \to \lambda_{\text{PF}}$ and normalization constants 
$C_{\psi,\phi}$ such that both $\bbox{\tilde{\psi}}$ and 
$\bbox{\tilde{\phi}}$ are normalized to one. We now proceed to
determine the scaling behavior of these normalization constants at
the critical point $\Lambda \equiv 0$. As the diagonal elements of the
$\bbox{\cal S}$-matrices vanish to leading order in $\Lambda$, 
the RT equations for $\bbox{\psi}$ and $\bbox{\phi}$ 
decouple at the critical point and can be considered separately. 
By taking successive \mbox{$\ast$-products} along the words $w_i$, the
{\em original}\/ eigenvector components may now be expressed
in terms of the {\em renormalized}\/ ones
\begin{equation}
\left(\begin{array}{@{}c@{}} \bbox{\psi}_{a_i^{},k}\\ 
\bbox{\phi}_{w_i^j,k+j+1}\end{array} \right) =
 \bbox{\cal S}_{a_i^{}|w_i^j}
\left(\begin{array}{@{}c@{}} \bbox{\phi}_{a_i^{},k+1}\\ 
\bbox{\psi}_{w_i^j,k+j}\end{array} \right)
\end{equation}
where $w_i^j$ denotes the $j$\/th letter of $w_i$ and
\begin{equation}
\bbox{\cal S}_{a_i^{}|w_i^j} = 
\bbox{\cal S}_{a_i^{}|w_i^1} \ast \bbox{\cal S}_{w_i^1|w_i^2} 
\ast \cdots \ast \bbox{\cal S}_{w_i^{j-1}|w_i^j} \; .
\end{equation}
Explicitly, we obtain the following relations
\begin{mathletters}\begin{eqnarray}
\bbox{\phi}_{a_i^{},k+1}^2 &=& 
(C_\phi^{-1}\bbox{\tilde{\phi}}_{a_i^{},m+1})^2 \; ,\\
\bbox{\phi}_{w_i^j,k+j+1}^2 &=& \frac{h_{a_i^{}}^2 h_{w_i^1}^2 \dots
h_{w_i^{j-1}}^2}{\varepsilon_{w_i^1}^2\varepsilon_{w_i^2}^2 \dots 
\varepsilon_{w_i^j}^2} (C_\phi^{-1}\bbox{\tilde{\phi}}_{a_i^{},m+1})^2 \; ,
\end{eqnarray}\end{mathletters}%
and
\begin{mathletters}\begin{eqnarray}
\bbox{\psi}_{a_i^{},k}^2 &=& 
(C_\psi^{-1}\bbox{\tilde{\psi}}_{a_i^{},m})^2 \; , \\
\bbox{\psi}_{w_i^j,k+j}^2 &=& 
\frac{\varepsilon_{a_i^{}}^2\varepsilon_{w_i^1}^2 \dots 
\varepsilon_{w_i^{j-1}}^2}{h_{a_i^{}}^2 h_{w_i^1}^2 \dots
h_{w_i^{j-1}}^2} (C_\psi^{-1}\bbox{\tilde{\psi}}_{a_i^{},m})^2 \; ,
\end{eqnarray}\end{mathletters}%
with the reduced couplings at their fixed point values.
In order to derive the renormalization factor of $C_\psi$ and 
$C_\phi$, we now perform partial sums of the 
squared vector entries, according to their additional labels $a_i$,
\begin{equation}
\bbox{\Psi}_{i}^{} = \sum_k \bbox{\psi}_{a_i,k}^2 \; , \quad
\bbox{\Phi}_{i}^{} = \sum_k \bbox{\phi}_{a_i,k}^2 \; .
\end{equation}
For these vectors $\bbox{\Psi}$ and $\bbox{\Phi}$,
the RT at the critical point yields a simple matrix form
\begin{equation}
C_\psi^{2} \bbox{\Psi} = \bbox{M}^{\psi} \bbox{\tilde{\Psi}}\; , \quad
C_\phi^{2} \bbox{\Phi} = \bbox{M}^{\phi} \bbox{\tilde{\Phi}}\; ,
\end{equation}
where
\begin{mathletters}\begin{eqnarray}
\bbox{M}_{ij}^{\psi} &=& 
\sum_{k=1}^{|w_i|} \delta_{w_i^k,a_j^{}}
\frac{h_{w_i^k}^2}{h_{a_i^{}}^2}
\frac{\varepsilon_{a_i^{}}^2}{\varepsilon_{w_i^k}^2}
\prod_{l=1}^k
\frac{\varepsilon_{w_i^l}^2}{h_{w_i^l}^2} + \delta_{a_i^{},a_j^{}} \; ,\\
\bbox{M}_{ij}^{\phi} &=& \sum_{k=1}^{|w_i|} \delta_{w_i^k,a_j^{}}
\frac{h_{a_i^{}}^2}{h_{w_i^k}^2} \prod_{l=1}^k
\frac{h_{w_i^l}^2}{\varepsilon_{w_i^l}^2} + \delta_{a_i^{},a_j^{}}\; .
\end{eqnarray}\end{mathletters}%
Since all components of the vectors $\bbox{\Psi}$, $\bbox{\Phi}$ and the
matrices $\bbox{M}^{\psi,\phi}$ are positive, the vectors converge to the
PF eigenvectors of $\bbox{M}^{\psi,\phi}$ under iteration of the RT.
We formally conclude, for the normalization constants,
\begin{equation}
C_\psi = \sqrt{\mu^{\psi}} \; , \quad C_\phi = \sqrt{\mu^{\phi}}\; ,
\label{cmu}
\end{equation}
where $\mu^{\psi,\phi}$ are the PF eigenvalues of 
$\bbox{M}^{\psi,\phi}$.
At this point, we can make contact to the RT of the lowest fermionic
excitations considered in Ref.~\onlinecite{HGB}. To this end, we
transform $\bbox{M}^{\psi,\phi}$ under conservation of the spectrum
according to 
$\bbox{M}^- = (\bbox{T}^\psi)^{-1} \bbox{M}^{\psi} \bbox{T}^\psi$ and 
$\bbox{M}^+ = (\bbox{T}^\phi)^{-1} \bbox{M}^{\phi} \bbox{T}^\phi$ with 
diagonal transformation matrices 
$\bbox{T}^\psi_{ij} = (\varepsilon_{a_i}/h_{a_i})^2 \delta_{i,j}$ and 
$\bbox{T}^\phi_{ij} = h_{a_i}^2 \delta_{i,j}$. We find
\begin{equation}\label{mpm}
\bbox{M}_{ij}^{\pm} = \sum_{k=1}^{|w_i|} \delta_{w_i^k,a_j^{}}
\prod_{l=1}^k
\left(\frac{h_{w_i^l}^2}{\varepsilon_{w_i^l}^2}\right)^{\pm 1} + 
\delta_{a_i^{},a_j^{}}\; .
\end{equation}
If $\mu^{-}\equiv\mu^{\psi}$ and $\mu^{+}\equiv\mu^{\phi}$ 
denote the leading eigenvalues of $\bbox{M}^\pm$, note that 
$\mu^{-}(\{\varepsilon_{a_i}/h_{a_i}\}) =
\mu^+(\{h_{a_i}/\varepsilon_{a_i}\})$. 

For substitution chains with bounded or marginal fluctuations of the
coupling constants, however, these are exactly the forms of the 
RT-matrices for the fermion frequencies [see Eqs.\ (3.21) and (3.25) in
Ref.~\onlinecite{HGB}]. For comparison, recall that one has
\begin{equation}
\prod_{l =1}^{|w_i|} \frac{h_{w_i^l}^2}{\varepsilon_{w_i^l}^2} = 1
\end{equation}
for $|\lambda_2| \le 1$.
The lowest fermion frequencies transform as 
$\tilde{\Lambda} = \sqrt{\mu^+ \mu^-} \Lambda$, resulting in the
scaling behavior \cite{HGB}
\begin{equation}
\Lambda_q = x_q N^{-z} \, , \quad z =
\frac{\log(\mu^+\mu^-)}{2\log \lambda_{\text{PF}}}\, .
\end{equation}

The extension of the RT to general substitution rules has been
described in Ref.~\onlinecite{HGB}; it can be performed along the same
lines here.  Generally, the substitution rule has to be redefined as a
function acting on {\em pairs} of letters, generating the same
chain. General $\bbox{\cal S}$-matrices $\bbox{\cal S}_{a_i|a_j}$
contain fields and asymmetry parameters, and act on vector components,
all carrying a double label $(\ldots )_{a_ia_j}$. The RT of the
eigenvector components can again be given as a matrix equation (of
dimension $n^2$) and identified with the RT of the fermionic
excitations as above.

\paragraph*{Results on critical scaling.}
The {\em surface magnetization}\/ $m_{s1}$ (on the {\em left}\/ surface) 
can be obtained from the large distance limit of the spin-correlation
function at the surface, leading to the expression \cite{SML,Pe}
\begin{equation}
m_{s1} = \langle 1| \sigma^x_1|0\rangle = \bbox{\phi}_1^{(1)} \; .
\end{equation}
Here, $|1\rangle = \bbox{\eta}_1^\dagger|0\rangle$ is the lowest
excited state which becomes degenerate with the ground state in the
ordered phase,\cite{SML} and the last equality is obtained upon
expressing $\sigma^x_1$ in terms of the fermions. Thus, $m_{s1}$ is
simply the first entry of the 
normalized eigenvector $\bbox{\phi}^{(1)}$ of the lowest 
fermionic excitation. Since
$\bbox{\tilde{\phi}}^{(1)}_1 = \sqrt{\mu^+} \bbox{\phi}^{(1)}_1$
[see Eqs.~(\ref{cc}) and (\ref{cmu}), and remember that 
$\mu^{+}\equiv\mu^{\phi}$], we obtain the surface 
magnetization exponent $\beta_{s1}$ through
\begin{equation}
m_{s1} \sim N^{-\beta_{s1}/\nu} \; , \quad \beta_{s1} = 
\frac{\log{\mu^+}}{2\log{\lambda_{\text{PF}}}}\; .
\end{equation}
Note that $\beta_{s1}$ is given by the scaling exponent here, since
the correlation length exponent (along the layers of the 2d classical
model) retains its unperturbed value $\nu = 1$.\cite{TIB,BBT} Using
a method of Ref.~\onlinecite{Pe}, the surface magnetization of several
substitution sequences was calculated before.\cite{TIB} However, it
seems impossible to extend this method to the general case. Recently,
the results of Ref.~\onlinecite{TIB} have also been rederived
\cite{ITKS} using real space renormalization.

For substitution chains with bounded fluctuations (and second largest
eigenvalue of the substitution matrix $|\lambda_2| < 1$), like e.g.\
the Fibonacci chain, all reduced couplings converge to the critical
value of the uniform chain $\varepsilon_{a_i}/h_{a_i} \to 1$ and we
find $\bbox{M}^+ = \bbox{M}^- = \bbox{M}_\varrho$, thus
$\mu^{+}=\lambda_{PF}$ and $\beta_{s1}$ takes its Onsager value
$\beta_{s1}=1/2$.

For two-letter substitution chains, which always can be generated by a
substitution rule of the form (\ref{subs}), \cite{HGB} an explicit result
can be obtained also for the marginal case ($\lambda_2 = 1$). Here, the 
PF eigenvalue of $\bbox{M}^\pm$ coincides with the trace of
these matrices,\cite{HGB} resulting in
\begin{equation}
\beta_{s1} = \frac{\log(\bbox{M}_{11}^+ + \bbox{M}_{22}^+ -1)}
{2\log \lambda_{\text{PF}}}\; 
\end{equation}
where $\bbox{M}_{11}^+$ and $\bbox{M}_{22}^+$ are given in Eq.~(\ref{mpm}). 
In general, $\beta_{s1}$ depends on the coupling constants through
$r<n$ independent parameters, where $r$ is the dimension of the 
(joined) eigenspace to eigenvalues $|\lambda| = 1$ of $\bbox{M}_\varrho$.

For substitution chains with fluctuations that diverge
with a power-law, the situation is slightly more complex.
Here, the reduced couplings scale as $\exp(\bbox{y}_i N^\omega)$; and
since $\bbox{y}$ always contains components of different sign, part
of them flow to infinity under the RT, while others converge to zero. 
For the RT of $C_{\phi,\psi}$, consider the behavior of
the entries of $\bbox{M}^{\pm}$ for {\em iterated}\/ substitution
steps $\varrho^n$. According to Eq.~(\ref{mpm}), all entries are 
polynomials of the reduced couplings.
Now, two cases have to be distinguished --- depending not only on
the substitution matrix, but on the detailed form of the substitution
itself. Generically, all entries are finally dominated by reduced
couplings that flow to infinity under the RT. This is always the case
if the cumulated deviation of any type of coupling from its mean is
unbounded in positive {\em and}\/ negative direction. In this case,
$m_{s1}$ will finally scale as $\exp(-cN^\omega)$, with $c$ depending
on the coupling constants. 

There are, however, special cases, where the cumulated deviation does
not diverge symmetrically, but, depending on the initial word (seed)
of the inflation, only in positive (or in negative) direction. As a
simple example, consider the two-letter substitution rule 
\begin{equation}
\varrho: \begin{array}{ccc} a& \to& aaab \\  b& \to& bbba \end{array} \;.
\end{equation}
Let $\sharp_{a,b}(n)$ be the number of $a,b$'s among the first 
$n$ letters of the limit chain; we find $1/2$ as the asymptotic frequency 
for both letters. Choosing, however, the single letter $a$ as the seed, the 
number $\sharp_a(n) - \sharp_b(n)$ is unbounded, but {\em positive}\/ 
for any $n$, thus the fluctuations are bounded from below for this chain.
For the corresponding matrices $\bbox{M}^\pm$, this means 
that, for given couplings, certain entries do not diverge, 
but finally tend to a finite value or to zero. Starting with the 
appropriate seed, this results
in a {\em finite}\/ surface magnetization $m_{s1}$ (on one side of the
chain), also at criticality. This explains the (unusual) behavior
that has been observed in Ref.~\onlinecite{IT94} for a special 
substitution rule.     

The {\em surface magnetization}\/ $m_{s2}$ on the {\em right}\/ 
surface follows, after an {\em inversion}\/ of the substitution rule to
\mbox{$\bar{\varrho}: a_i \to \bar{w}_i a_i$},
through the same procedure as described above. The RT for the 
vector components reads
\begin{equation}
\left(\begin{array}{@{}c@{}} \bbox{\psi}_{w_i^j,k-j}\\
\bbox{\phi}_{a_i^{},k+1}\end{array} \right) =
 \bbox{\cal S}_{w_i^j|a_i^{}}
\left(\begin{array}{@{}c@{}} \bbox{\phi}_{w_i^j,k-j+1}\\ 
\bbox{\psi}_{a_i^{},k}\end{array} \right) \; ,
\end{equation}
and we deduce the same relations as above {\em with all reduced
couplings inverted}\/, i.e.,  
$\varepsilon_{a_i}/h_{a_i} \to h_{a_i}/\varepsilon_{a_i}$.
Thus 
\begin{equation}
\beta_{s2}\bigl(\bigl\{\frac{\varepsilon_{a_i}^{}}{h_{a_i}}\bigr\}\bigr)=
\beta_{s1}\bigl(\bigl\{\frac{h_{a_i}}{\varepsilon_{a_i}^{}}\bigr\}\bigr)= 
\frac{\log \mu^-}{2\log\lambda_{\text{PF}}} \; .
\end{equation}
This generalizes a symmetry in scaling of $m_{s1}$ and $m_{s2}$
\cite{BBT,ITKS} also to substitution chains 
where this symmetry is not obvious from the chain geometry, and to
general $n$-letter substitutions. 
We also conclude, for arbitrary $n$-letter substitution chains
with marginal (or bounded) fluctuations of the coupling constants, that
\begin{equation}
z = \beta_{s1} + \beta_{s2} 
\end{equation}
as conjectured in Ref.~\onlinecite{BBT} and argued-for 
in Ref.~\onlinecite{ITKS}. 

The {\em surface energy density}\/ can be obtained from the two lowest 
fermionic excitations as follows \cite{BBT}
\begin{equation}
e_{s1} \sim (\Lambda_2 - \Lambda_1)\, 
\bbox{\phi}_1^{(1)}\bbox{\phi}_1^{(2)} \; .
\end{equation}
Hence, we find the scaling behavior $e_{s1,2}\sim N^{-z-2\beta_{s1,2}}$ 
conjectured in Ref.~\onlinecite{BBT} on the basis of numerical results.

The {\em bulk energy density}\/ on site $j$ follows from the matrix 
element \cite{BBT}
\begin{equation}
e_k = \left|\bbox{\psi}_k^{(1)} \bbox{\phi}_k^{(2)} -
\bbox{\psi}_k^{(2)}\bbox{\phi}_k^{(1)}\right| \; .
\end{equation}
Thus, $\tilde{e}_m = \sqrt{\mu^+\mu^-} e_k$ and $e_{N/k} \sim
N^{-z}$, as had also been conjectured in Ref.~\onlinecite{BBT}.  

\paragraph*{Conclusions.}
We have demonstrated how the bulk and surface critical behavior of
aperiodic Ising quantum chains (or, equivalently, of aperiodically
layered, two-dimensional classical Ising systems) obtained from
general substitution rules can be derived by exact real space
renormalization. The results corroborate a number of conjectures on
scaling laws for these systems and generalize previous results
obtained for special examples.  An exact calculation of the bulk
magnetization seems to be more difficult and remains an open problem.

\paragraph*{Acknowledgement.}
The authors thank M.~Baake for discussions and helpful comments.


\begin{references}
\bibitem{TIB}
L.~Turban, F.~Igl\'{o}i and B.~Berche,
       Phys.\ Rev.\ B {\bf 49}, 12695 (1994).
\bibitem{IT94}
F.~Igl\'{o}i and L.~Turban,
        Europhys.\ Lett.\ {\bf 27}, 91 (1994).
\bibitem{BBT}
B.~Berche, P.-E.~Berche and L.~Turban,
        J.\ Phys.\ France I {\bf 6}, 621 (1996).
\bibitem{IT}
F.~Igl\'{o}i and L.~Turban,
        Phys.\ Rev.\ Lett.\ {\bf 77}, 1206 (1996).
\bibitem{HGB}
J.~Hermisson, U.~Grimm and M.~Baake,
       J.\ Phys.\ A {\bf 30}, 7315 (1997).
\bibitem{ITKS}
F.~Igl\'{o}i, L.~Turban, D.~Karevski and F.~Szalma,
       Phys.\ Rev.\ B {\bf 56}, 11031 (1997).   
\bibitem{GB96}
U.~Grimm and M.~Baake,
       in: {\it The Mathematics of Long-Range Aperiodic Order\/},
       edited by R.\thinspace V.~Moody
       (Kluwer, Dordrecht, 1997), pp.~199--237.
\bibitem{L}
J.\thinspace M.~Luck, 
        J.\ Stat.\ Phys.\ {\bf 72}, 417 (1993);
        Europhys.\ Lett.\ {\bf 24}, 359 (1993).
\bibitem{LSM}
E.\thinspace H.~Lieb, T.~Schultz and D.~Mattis,
        Ann.\ Phys.\ (NY) {\bf 16}, 407 (1961).
\bibitem{RED}
R.~Redheffer, 
       in: {\it Modern Mathematics for the Engineer},
       ed. E.~F.~Beckenbach, McGraw Hill, New York (1961), pp.~282--337.
\bibitem{SML}
T.~Schultz, D.~Mattis and E.\thinspace H.~Lieb,
        Rev.\ Mod.\ Phys.\ {\bf 36}, 856 (1964).
\bibitem{Pe}
I.~Peschel,
        Phys.\ Rev.\ B {\bf 30}, 6783 (1984).
\end{references}
\end{document}